\documentclass[12pt]{article}

\usepackage{epsfig}
\usepackage{epsf}

\def\mathbf{\vec}

\def\ca{\c{c}\~{a}}

\def\ii{\'{\i}}

\begin{document}

\centerline{\bf {\large The role of hidden ambiguities in the }}
\centerline{\bf {\large Linear Sigma Model with fermions }}
\vspace{1cm}

\centerline{Brigitte Hiller$^{(1)}$, A. L. Mota$^{(2),(5)}$, M.C. Nemes$^{(1),(3)}$,}
\centerline{Alexander A. Osipov$^{(1),(4)}$ and Marcos  Sampaio$^{(3)}$}
\vspace{1cm}

{\small {\it (1) Centro de F\'{\i}sica Te\'{o}rica, Departamento de
         F\'{\i}sica da Universidade de Coimbra, 3004-516 Coimbra, Portugal
\vspace{.5cm}

(2) Departamento de Ci\^encias Naturais, Universidade Federal de S\~ao Jo\~ao del Rei, S\~ao Jo\~ao del Rei, MG, Brazil
\vspace{.5cm}

(3) Departamento de F\ii sica, Instituto de Ci\^encias Exactas, Universidade
Federal de Minas Gerais, BH, {\it CEP} 30161-970, MG, Brazil \vspace{.5cm}

(4) Joint Institute for Nuclear Research, Laboratory of Nuclear Problems, 141980 Dubna,
        Moscow Region, Russia \vspace{.5cm}

(5) Departamento de Fisica Moderna, Faculdad de Ciencias, Universidad de Granada,
    Granada, Spain}}
\vspace{1cm}

\centerline{{\bf Abstract}}
\vspace{.5cm}

{\small The $U_L(3)\times U_R(3)$ Linear Sigma Model (LSM) with quark degrees of freedom is used to show that radiative
corrections generate undetermined finite contributions. Their origin is related to surface terms which are differences
between divergent integrals with the same degree of divergence. The technique used to detect these ambiguities is an
implicit regularization on basic divergent integrals that do not depend on external momenta. We show that
such contributions are absorbed by renormalization or fixed by symmetry requirements. The general expression for surface terms is derived. Renormalization group coefficients
are calculated, as well as  relevant observables for this model, such as $f_{\pi}$, $f_k$ and the pion and kaon form factors.}
\vspace{1cm}
\pagebreak

{\bf I. Introduction}
\vspace{1cm}

In dealing with  ultraviolet infinities which stem
from the semi-classical expansion in a quantum field
theory, the renormalization program plays an essential
role to fit the experimental data which
the model is meant to describe.   In regularizing
and subtracting a divergence  there appears a hidden parameter which plays
the role of renormalization group scale in renormalizable models.   The imposition that the Green functions do not
depend on how the infinities are subtracted
from the amplitudes can be formulated by demanding that the Green functions are
independent of such parameter. This gives rise to
the running couplings, quantum mechanical mass generation, etc.

Recently Jackiw \cite{Jackiw:2000} raised an interesting
matter concerning arbitrary parameters which appear in perturbative calculations
in certain field theoretical models. Such arbitrary (regularization dependent)
parameters show up as differences between divergent
integrals of the same superficial degree of
divergence. Usually physicists do not like to have
free parameters in their theoretical models. The
prescription to deal with such arbitrary quantities is
threefold: 1) check whether they may be considered as
finite counterterms (as it is usual in renormalizable
models); 2) check whether symmetry constraints (e.g.
supersymmetry, gauge symmetry, etc.) may require that
such parameters assume a definite value and 3) see if
as a genuine free parameter, phenomenology can make
use of it to fit the data in the underlying model.  Of
course such parameters are expected to play an
essential role in finite theories and effective
models.

A well known example where a finite ambiguity plays an
essential role is in the description of the
Adler-Bardeen-Bell-Jackiw anomaly  \cite{Bell:1969},
\cite{Adler:1969}. In this
case the finite constant is a manifestation of an
important symmetry breaking at the level of
perturbative quantum corrections. A democratic
description of the anomaly between the vector and
axial sectors is achieved when we allow a
regularization dependent parameter to be arbitrary
\cite{anom}.

The ideal arena to implement such ideas in the momentum space is the Implicit
Regularization (IR) scheme \cite{orimar:1999}, which has  already been
successfully applied to several examples
\cite{baeta1:2001},\cite{anom},\cite{gobira:2004},\cite{carneiro:2004},\cite{souza:2005}.
The idea behind IR is that one

(a)  separates external momentum independent divergencies from finite contributions, without the need of an explicit
regulator;

(b)  leaves their divergent content in the form of basic integrals until the end of the calculation;

(c)  does not modify the underlying theory, preserving its symmetries and
space-time dimension.

The purpose of the present contribution is the investigation in the LSM \cite{Levy:1960} of
the role played by ambiguities
in the context of the above discussed ideas. We show that radiative
corrections yield together with well known
infinities also finite undetermined contributions.
In this work we will follow Jackiw's suggestion and leave the
arbitrary quantities open until the end of the calculations. Then, we
will show that these undetermined contributions can be fixed
by symmetry relations or absorbed by renormalization, in agreement
with the ideas presented in \cite{Bonneau2000}. From the point of view of
the current  phenomenology, because the present Lagrangian does not contain
the important 't Hooft determinantal interaction
\cite{Hooft:1976}, we restrict the fits to the pion and kaon sectors, which
are not crucially dependent on it.

The work is
organized as follows: in section 2. the method is presented for the case of
amplitudes containing different masses in the fermion propagators and we show
the origin of ambiguities. The renormalization group scale is also introduced
within IR. In section 3. the model is renormalized and some ambiguities are
shown to be absorbed by renormalization. Moreover the $\beta$-function is
explicitly calculated and the renormalization group equation for the pion
coupling constant is solved in section 3A. The position of the well-known Landau
pole is identified  and coincides with the position of the pole in the large
euclidean momentum region of the pion propagator. In section 4. we derive the
expressions for the weak decay constants $f_\pi$ and $f_{\kappa}$, as well as the
electromagnetic form factor for the kaon. Numerical results are given in section
5., and conclusions are given in section 6. \vspace{1cm}

{\bf 2. Treatment of divergent integrals, ambiguities and renormalization group scale}
\vspace{1cm}

In the present section we illustrate the relevant technical details of the renormalization procedure
we use by working out explicitly a one-loop Feynman amplitude.

Consider the pseudoscalar amplitude with two quark masses $m_u$ and $m_s$
\begin{equation}
\label{PP}
  {\Pi}^{PP}=i\int_{\Lambda} \frac{d^4 k}{(2 \pi)^4} \mbox{Tr}\left(
  \gamma_5 \frac{1}{\not k - m_u}\gamma_5
  \frac{1}{\not k -  \not p - m_s}\right),
\end{equation}
where the symbol $\Lambda$ stands for a regulator which needs not be explicitated, but is necessary to give a meaning to
${\Pi}^{PP}$. One is then allowed to algebraically manipulate the integrand. We do it in such a way that the divergencies
appear as integrals and are separated from the finite (external momentum dependent) contribution to Eq. (\ref{PP}). After
taking the Dirac trace, we use the following algebraic identity

\begin{equation}
\label{reduction}
  \frac{1}{(k-p)^2-m^2}=\frac{1}{k^2-m^2}-\frac{p^2-2p.k}{[(p-k)^2-m^2](k^2-m^2)}
\end{equation}
at the level of the integrand. Note that it allows one to confine the external momentum dependence in more
convergent integrals. This relation should be used recursively until the finite part is completely separated from divergent integrals.
We get
\begin{eqnarray}
\label{PP1}
  {\Pi}^{PP}&\!\!\! =\!\!\!&-2\left\{\int \frac{d^4 k}{(2 \pi)^4}
  \frac{i}{k^2-m_u^2} 
  +\int\frac{d^4 k}{(2 \pi)^4}\frac{i}{(k-p)^2-m_s^2}
  \right.\nonumber \\
  &\!\!\! +\!\!\!&\left. [(m_s-m_u)^2-p^2] \int \frac{d^4 k}{(2 \pi)^4}
      \frac{i}{(k^2-m_u^2)[(k-p)^2-m_s^2]}\right\}.
\end{eqnarray}

The first integral on the RHS is what we call a basic quadratic divergence
\begin{equation}
\label{Iquad}
   I_q(m^2)=\int \frac{d^4 k}{(2 \pi)^4}\frac{i}{k^2-m^2}.
\end{equation}

The second integral on the RHS is also a quadratic divergence, but it still possesses an external momentum dependence.
If one uses Eq. (\ref{reduction}), one sees that an arbitrariness emerges
\begin{equation}
\label{arb}
  T(p^2,m_i)= \int \frac{d^4 k}{(2 \pi)^4}\frac{i}{(k-p)^2-m_s^2}
  = I_q(m_s^2)+p^{\mu}p^{\nu}\Delta_{\mu\nu},
\end{equation}
where
\begin{equation}
\label{alpha}
   \Delta_{\mu\nu}=i\int \frac{d^4 k}{(2 \pi)^4}
   \frac{4 k_\mu k_\nu}{(k^2-m_s^2)^3}-i\int
   \frac{d^4 k}{(2 \pi)^4}\frac{g_{\mu\nu}}{(k^2-m_s^2)^2}
   =\alpha g_{\mu\nu}, 
\end{equation}
and $\alpha =\mbox{const}$, i.e. the difference between two logarithmically divergent integrals. Eq. (\ref{alpha}) is a surface term, a
particular case of the general expression for surface terms (see derivation in Appendix )
\begin{equation}
   S^{(A-2)}_{\mu_{1}\mu_{2}\ldots \mu_{2n}}=i
   \int \frac{d^{2\omega}p}{(2\pi )^4}
   \left[
   \frac{g_{\mu_{1}\mu_{2}\ldots \mu_{2n}}
    }{(p^2-m^2)^{A-n-1}}-
   \frac{2^n\Gamma (A-1)}{\Gamma (A-n-1)}
   \frac{p_{\mu_{1}} p_{\mu_{2}}\ldots p_{\mu_{2n}}
   }{(p^2-m^2)^{A-1}}\right]
\end{equation}
\noindent where $\omega$ is a continuous dimension, $\mu_i$ is a Lorentz index and
$g_{\mu_{1}\mu_{2}\ldots \mu_{2n}}$ is the standard notation for symmetrized products of the $g_{\mu\nu}$ tensor. It is
clear that at $\omega=2$, $A=4$ and $n=1$ we reproduce Eq. (\ref{alpha}).
In dimensional regularization or Pauli-Villars, following the usual prescriptions to preserve gauge invariance, one would get
zero for $\Delta_{\mu\nu}$. However for other regularization prescriptions one
may expect a dimensionless number. Following Jackiw's suggestion we leave this
arbitrary number open until the end of the calculations and only then see if it
must be fixed by Ward identities. In QED and QCD gauge invariance forces it to
be zero \cite{baeta1:2001}. In the anomalous pion decay, on the other hand, it must  be
finite \cite{anom}.

The third term in Eq. (\ref{PP1}) is also logarithmically divergent. In the
spirit of Implicit Regularization the integral can be displayed as
\begin{eqnarray}
\label{finite}
   \int\! \frac{d^4 k}{(2 \pi)^4}\frac{i}{(k^2-m_u^2)[(k-p)^2-m_s^2]}
   &\!\!\! =\!\!\!& \int\! \frac{d^4 k}{(2 \pi)^4}\frac{i}{(k^2-m_u^2)^2}
    + F(m_u^2,m_s^2,p^2;m_u^2)\nonumber \\
   &\!\!\! =\!\!\!& \int\! \frac{d^4 k}{(2 \pi)^4}\frac{i}{(k^2-m_s^2)^2}+ 
   F(m_u^2,m_s^2,p^2;m_s^2)\nonumber \\
   &\!\!\! =\!\!\!& \int\! \frac{d^4 k}{(2 \pi)^4}\frac{i}{(k^2-\zeta^2)^2}+ 
   F(m_u^2,m_s^2,p^2;\zeta^2)\nonumber \\
\end{eqnarray}
\noindent where again we separated another basic divergent integral
\begin{equation}
\label{Ilog}
  I_{log}(m^2)=\int \frac{d^4 k}{(2 \pi)^4}\frac{i}{(k^2-m^2)^2}\ ,
\end{equation}
\noindent and a finite contribution
\begin{equation}
\label{finite1}
F(m_u^2,m_s^2,p^2;\zeta^2)=\frac{1}{(4\pi)^2}\int_0^1 dz {\it ln} (\frac{p^2 z (1-z)+(m_u^2-m_s^2) z -m_u^2}{-\zeta^2}).
\end{equation}

In this expression the last argument corresponds to a squared mass in the denominator of the logarithm and plays the role of a scale.
This comes from the property of the logarithmic divergence,
\begin{equation}
\label{scale}
  I_{log}(m^2)=I_{log}(\zeta^2) +\frac{1}{(4 \pi)^2}ln(\frac{m^2}{\zeta^2}),
\end{equation}
\noindent and its intimate connection with the finite part of the amplitude.
The arbitrary scale $\zeta^2$ represents in this method the renormalization group scale.

Finally we obtain that
\begin{eqnarray}
\label{PPf}
  {\Pi}^{PP}&=&2\{I_q(m_u^2) +   I_q(m_s^2) + p^2 \alpha
      \nonumber \\
   &+&  [(m_u-m_s)^2-p^2] (I_{log}(\zeta^2) + F(m_u^2,m_s^2,p^2;\zeta^2))\}.
\end{eqnarray}

Analogously one finds for the scalar amplitude
\begin{equation}
\label{SS}
  {\Pi}^{SS}=i\int_{\Lambda} \frac{d^4 k}{(2 \pi)^4} \mbox{Tr} 
  \left( \frac{1}{\not k - m_u} \frac{1}{\not k\ -\not p-m_s}\right),
\end{equation}

the following relation
\begin{eqnarray}
\label{PPs}
  {\Pi}^{SS}&=&2\{I_q(m_u^2) +   I_q(m_s^2) + p^2 \alpha
      \nonumber \\
   &+& [(m_u+m_s)^2-p^2] (I_{log}(\zeta^2) + F(m_u^2,m_s^2,p^2;\zeta^2))\}.
\end{eqnarray}

\vspace{1cm}

{\bf 3. The model}
\vspace{1cm}

We start by considering the following generating functional
\begin{eqnarray}
\label{path}
   Z&=&\int \prod_a {\cal D} \sigma_{0a}{\cal D} \pi_{0a} {\cal D}q 
      {\cal D}{\bar q}\ \mbox{exp}(iS(\bar{q},q,\sigma_0,\pi_0))
\end{eqnarray}
with the action $S(\bar{q},q,\sigma_0,\pi_0)$ of a $U(3)\times U(3)$ 
linear sigma-model including fermionic degrees of freedom
\begin{eqnarray}
\label{action}
  S(\bar{q},q,\sigma_0,\pi_0)&\!\!\! =\!\!\!&\int d^4x 
  \left[ {\cal L}_q -\frac{\mu_0^2}{4}\ \mbox{tr} (BB^{\dagger})
  -\frac{\lambda_{0q}}{2}\ \mbox{tr} [(BB^{\dagger})^2]
  +\frac{1}{2 g_0}\ \mbox{tr}(c_0\sigma_0)
  \right. \nonumber \\
  &\!\!\! +\!\!\!&\left. \frac{f_0^2}{4}\ \mbox{tr}
  (\partial_\mu \sigma_0 \partial^{\mu} \sigma_0 
  +\partial_\mu \pi_0 \partial^{\mu} \pi_0)\right],
\end{eqnarray}
\begin{equation}
\label{Lq}
   {\cal L}_{q}=\bar q [i\!\not \!\partial
   - g_0 (\sigma_0 + i\gamma_5 \pi_0)]q.
\end{equation}

\noindent Here $B=\sigma_0 + i\pi_0$, $B^{\dagger}=\sigma_0 - i\pi_0$, used in the definition of the quadratic and quartic invariants in mesonic fields  \cite{Gasiorowicz:1969}
\begin{eqnarray}
\label{invariant}
   \mbox{tr}(BB^{\dagger})&\!\!\! =\!\!\!&\mbox{tr}(\sigma_0^2 +\pi_0^2)
   \nonumber \\
   \mbox{tr}[(BB^{\dagger})^2]&\!\!\! =\!\!\!&\mbox{tr}
   \{(\sigma_0^2 +\pi_0^2)^2-[\sigma_0,\pi_0]^2\},
\end{eqnarray}
\noindent with the scalar and pseudoscalar fields $\sigma_0=\lambda_a \sigma_0^a$ and $\pi_0=\lambda_a \pi_0^a$ containing
the standard $U(3)$ matrices $\lambda_a$ (a=0,1...8) in flavor space.  The subscript 0 stands for bare quantities,
${\cal L}_{q}$ describes the coupling of quark fields $q,\bar{q}$ to the mesons, $g_0$ is dimensionless. 

The symmetry transformation properties of the fermionic fields are
\begin{eqnarray}
   \delta q&\!\!\! =\!\!\!&i(\alpha' +\gamma_5 \beta')q, \nonumber \\
   \delta{\bar q}&\!\!\! =\!\!\!& -i{\bar q} (\alpha'-\gamma_5 \beta')
\end{eqnarray}
\noindent and for the mesonic fields
\begin{eqnarray}
\label{trans}
   \delta \sigma_0 &\!\!\! =\!\!\!& i [\alpha',\sigma_0]+\{\beta',\pi_0\}
   \nonumber \\
   \delta \pi_0&\!\!\! =\!\!\!&i [\alpha',\pi_0]-\{\beta',\sigma_0\}
\end{eqnarray}

\noindent where the parameters of the infinitesimal global transformations $\alpha'$ and $\beta'$ are Hermitian flavour matrices.
The explicit symmetry breaking piece is introduced via the linear term proportional to $c_0 \sigma_0$ with a diagonal
matrix-valued strength $c_0=diag\{c_{0u},c_{0u},c_{0s}\} $. Without the quartic interaction and kinetic terms for the mesons,
proportional to $f_0^2$ and $\lambda_{0q}$ respectively, the action corresponds to a  Nambu-Jona-Lasinio (NJL) type lagrangian \cite{Nambu:1961}
in semi-bosonized form; this is achieved by the method of auxiliary fields \cite{Eguchi:1975}, which has been widely used in various
extensions of the original NJL lagrangian, see e.g. \cite{Reinhardt:1988}-\cite{Osipov:2001}. We relate the $c_0$ to the
diagonal current quark mass matrix $\hat {m}_0$ through the choice $c_0=\mu^2_0 \hat {m}_0$. We restrict our study to the case $\hat{m}_u=\hat{m}_d\ne \hat{m}_s$, which breaks the unitary $SU(3)$ symmetry down to the subgroup $SU(2)_I\times U(1)_Y$ (isospin-hypercharge).
Upon the exact integration over the quark variables in the path integral Eq. (\ref{path}), one obtains for the real part of the effective
 action associated with the fermions (we do not consider here anomalous processes),
\begin{equation}
\label{qi}
   \int {\cal D}q {\cal D}{\bar q} \mbox{ exp}(i\int d^4x {\cal L}_q) 
   \to \mbox{exp} (\ln |\det D_E|) =\mbox{ exp}\left(\frac{1}{2}\mbox{Tr}
   \ln (D_E^{\dagger}D_E)\right)
\end{equation}
\noindent where we use the strictly positive unbounded Hermitian second order elliptic operator $D_E^{\dagger}D_E$, which is chiral and gauge covariant \cite{Schwinger:1951}-\cite{Ball:1989} and where $\mbox{Tr}$ designates functional trace, $D_E$ stands for the euclidean Dirac operator
\begin{eqnarray}
\label{D}
   D_E&\!\!\! =\!\!\!&i\gamma_\mu \partial_\mu 
              - g_0(\sigma_0 +i \gamma_5 \pi_0),
              \nonumber \\
   D^{\dagger}_E&\!\!\! =\!\!\!&-i\gamma_\mu \partial_\mu 
              - g_0(\sigma_0 -i \gamma_5 \pi_0).
\end{eqnarray}
Since the scalar fields  possess  non-vanishing vacuum expectation values, we perform the shift $\sigma_0\rightarrow
\frac{m}{g_0}+\sigma_0$, where m represents the finite constituent quark mass
matrix and obtain after this shift
\begin{equation}
D_E^{\dagger}D_E=m^2-\partial^2 + Y,
\end{equation}
\noindent with the background mesonic fields \cite{Osipov:2004}

\begin{equation}
   Y=i g_0\gamma_\mu(\partial_\mu\sigma_0+i\gamma_5\partial_\mu\pi_0)
   + g_0^2\left(\sigma_0^2 +\{{\frac{m}{g_0},\sigma_0}\}
   +\pi_0^2 + i\gamma_5 [\sigma_0+\frac{m}{g_0},\pi_0]\right),
\end{equation} 

\noindent leading to the expansion
\begin{eqnarray}
\label{DD}
   \mbox{Tr} (\ln D_E^{\dagger}D_E)&\!\!\! =\!\!\!&
   \sum_{n=1}^{\infty} \frac{1}{n}\mbox{Tr}\{(-\partial^2+m^2)
   [1+(-\partial^2+m^2)^{-1}[i g_0 \gamma_\mu ( \partial_\mu 
   \sigma_0+i\gamma_5 \partial_{\mu}\pi_0)
   \nonumber\\
   &&+g_0^2(\sigma_0^2+\{\sigma_0,m/g_0\}+\pi_0^2
     +i\gamma_5[\pi_0,\sigma_0+m/g_0])]]\}^n.
\end{eqnarray}

From now on it is convenient to use the following representation  of the scalar and pseudoscalar fields
\begin{equation}
\label{fields}
   \frac{\lambda_a\sigma_{a}}{\sqrt{2}}=\left(
          \begin{array}{ccc}
          \frac{\sigma_u}{\sqrt{2}}&\sigma^+& \delta^+ \\
          \sigma^-& \frac{\sigma_d}{\sqrt{2}}& \delta^0 \\
          \delta^-&\bar \delta^0&\frac{\sigma_s}{\sqrt{2}}
          \end{array} \right),\qquad
   \frac{\lambda_a\pi_{a}}{\sqrt{2}}=\left(
          \begin{array}{ccc}
          \frac{\phi_u}{\sqrt{2}}&\pi^+&K^+ \\
          \pi^-&\frac{\phi_d}{\sqrt{2}}&K^0 \\
          K^-&\bar K^0&\frac{\phi_s}{\sqrt{2}}
          \end{array} \right).
\end{equation}

The gap equation will be obtained by considering the $n=1$ term in the expansion Eq. (\ref{DD}) and the linear
contributions in $\sigma_0$ from the remaining terms of Eq. (\ref{action}) after the mass shift. We get

\begin{equation}
\label{gap}
\frac{\mu_0^2}{g_0^2}m_i-\frac{c_{0i}}{g_0^2}-8N_c m_i I_q(m_i)+2\frac{\lambda_{0q}}{g_0^4}m_i^3=0,
\end{equation}

\noindent where $\{i=u,s\}$.

The $n=2$ term in the expansion contains all the other divergent contributions which go up to four-point functions.
In the calculations there appear the two basic divergent integrals $I_q$ and $I_{log}$, Eqs. (\ref{Iquad}) and (\ref{Ilog}), as
well as the difference between two logarithmically divergent integrals, $\alpha$, Eq. (\ref{alpha}). Expressions (\ref{PPf}) and
 (\ref{PPs}), after properly taking into account $g_0$ and trace factors in color ($N_c$) and Dirac spaces, are the amplitudes,
 continued to Minkowski space, resulting from terms quadratic in the fields of (\ref{DD}).
\vspace{0.5cm}

In the following we need the wave function renormalizations, which are readily obtained as the coefficients of the $p^2$ terms
in all expressions quadratic in the fields, stemming from the kinetic terms proportional to $f_0^2$ and from the amplitudes (\ref{PPf}),
(\ref{PPs})
\begin{eqnarray}
\label{Zm1s}
Z^{-1}_{\sigma}(m_i,m_j;\zeta^2)&=&f_0^2-4N_c g_0^2\{I_{log}(\zeta^2)-\alpha+F(m_i,m_j,0;\zeta^2)
\nonumber\\
                        &-&(m_i+m_j)^2 F'(m_i,m_j,0;\zeta^2)\},
\end{eqnarray}
\begin{eqnarray}
\label{Zm1p}
Z^{-1}_{\pi}(m_i,m_j;\zeta^2)&=&f_0^2-4N_c g_0^2\{I_{log}(\zeta^2)-\alpha+F(m_i,m_j,0;\zeta^2)
\nonumber\\
                        &-&(m_i-m_j)^2 F'(m_i,m_j,0;\zeta^2)\}.
\end{eqnarray}
\noindent where $F'$ represents the derivative with respect to square external
momentum of $F$, taken at $p^2=0$. The retained finite terms
$F(m_i,m_j,0;\zeta^2)$ and $F'(m_i,m_j,0;\zeta^2)$ lead to different wave
function renormalizations for the different members of the pseudoscalar and
scalar nonets; the higher contributions to the expansion of
$F(m_i,m_j,p^2;\zeta^2)$ will be absorbed in the renormalized propagators of the
mesons (see Eq. (\ref{propa}) below).

The renormalized coupling constants ${g_{\sigma}(m_i,m_j)}$ and ${g_{\pi}(m_i,m_j)}$ are then obtained as
\begin{equation}
\label{gsi}
\frac{g_0^2}{g^2_{\sigma}(m_i,m_j;\zeta^2)}= Z^{-1}_{\sigma}(m_i,m_j;\zeta^2)
\end{equation}
\begin{equation}
\label{gp}
\frac{g_0^2}{g^2_{\pi}(m_i,m_j;\zeta^2)}= Z^{-1}_{\pi}(m_i,m_j;\zeta^2),
\end{equation}

and the renormalized masses $\mu_{\sigma,\pi}(m_i,m_j;\zeta^2)$ become
\begin{eqnarray}
\label{Ms}
\frac{\mu^2_{\sigma,\pi}(m_i,m_j;\zeta^2)}{Z_{\sigma,\pi}(m_i,m_j;\zeta^2)}&=&M^2_{\sigma,\pi}(m_i,m_j;\zeta^2)
\nonumber\\
&=&\mu_0^2-4N_c g_0^2((I_q(m_i^2)+I_q(m_j^2))
\nonumber\\
&-&4N_cg_0^2(m_i\pm m_j)^2(I_{log}(\zeta^2)+F(m_i,m_j,0;\zeta^2))
\nonumber\\
&+&2 \frac{\lambda_{0q}}{g_0^2}(m_i^2+m_j^2 \pm m_im_j),
\end{eqnarray}
where $M^2_{\sigma}$ goes with the plus signs, and the renormalized quartic coupling
\begin{equation}
\frac{\lambda_q}{g_\pi^2(m_u,m_u;\zeta^2)}=\frac{\lambda_{0q}}{g_0^4}-4N_cI_{log}(\zeta^2).
\end{equation}

We have retained in the renormalization of the masses the finite terms $F(m_i,m_j,0;\zeta^2)$.
Note that all ambiguities $\alpha$ appear in the field renormalization coefficients (\ref{Zm1s}-\ref{Zm1p}). In principle they
could assume different values for the different processes. Chiral symmetry restricts them to have the same value, as we shall
see below.
Using relations (\ref{gp}) and (\ref{Ms}) one obtains that the kaon and pion renormalization constants are related as
(where we have identified $g_\pi=g_{\pi}(m_u,m_u;\zeta^2)$ and $g_\kappa=g_{\pi}(m_u,m_s;\zeta^2)$)
\begin{eqnarray}
\label{gkgp}
   \frac{1}{g^2_\kappa}&\!\!\! =\!\!\!&\frac{1}{g^2_\pi}
   -4N_c[F(m_u,m_s,0;\zeta^2)-F(m_u,m_u,0;\zeta^2)
   \nonumber \\
   &\!\!\! -\!\!\!&(m_u-m_s)^2 F'(m_u,m_s,0;\zeta^2)] \\
\label{ukup}
   \frac{\mu^2_\kappa}{g^2_\kappa}&\!\!\! =\!\!\!&\frac{\mu^2_\pi}{g^2_\pi}
   +2 m_s(m_s-m_u) \left[\frac{\lambda_q}{g^2_\pi} 
   -4N_cF(m_u,m_s,0;\zeta^2)\right]
\end{eqnarray}
\begin{equation}
   F(m_u,m_s,0;\zeta^2)=-\frac{1}{(4\pi)^2} \left[ 
   \frac{m_s^2}{m_s^2-m_u^2} \ln\left( \frac{m_s^2}{m_u^2}\right)
   +\ln\left(\frac{m_u^2}{\zeta^2}\right)-1\right].
\label{fmums0}
\end{equation} 
With these definitions we obtain the effective action
\begin{equation}
\label{Sterms}
S=S_{kin}+S_{mass}+S_{int},
\end{equation}
where $S_{kin},S_{mass},S_{int}$ are the actions related to kinetic, mass and interaction terms of the mesons
(in the equations for $S_{kin}, S_{mass}$ below we omit the index 0 in the bare fields in order not to clutter the notation).
\begin{eqnarray}
   S_{kin}&\!\!\! =\!\!\!&\int d^4x \left\{ 
   Z^{-1}_{\sigma}(m_u,m_u)(\partial_\mu\sigma_u\partial^\mu\sigma_u 
   +\partial_\mu \sigma_d \partial^\mu \sigma_d +\partial_\mu \sigma^+ 
   \partial^\mu \sigma^-) 
   \right.\nonumber\\
   &\!\!\!+\!\!\!& Z^{-1}_{\sigma}(m_s,m_s)\partial_\mu\sigma_s 
   \partial^\mu \sigma_s
   \nonumber\\
   &\!\!\! +\!\!\!& Z^{-1}_{\sigma}(m_u,m_s)(\partial_\mu\delta^+ 
   \partial^\mu \delta^-+\partial_\mu \delta^0 \partial^\mu \bar \delta^0) 
   \nonumber\\
   &\!\!\! +\!\!\!& Z^{-1}_{\pi}(m_u,m_u)(\partial_\mu\phi_u\partial^\mu 
   \phi_u +\partial_\mu \phi_d \partial^\mu \phi_d+\partial_\mu \pi^+
   \partial^\mu \pi^-)
   \nonumber\\
   &\!\!\! +\!\!\!& Z^{-1}_{\pi}(m_s,m_s)\partial_\mu \phi_s
   \partial^\mu \phi_s \nonumber\\
   &\!\!\! +\!\!\!&\left.  Z^{-1}_{\pi}(m_u,m_s)(\partial_\mu K^+ 
   \partial^\mu K^-+\partial_\mu K^0 \partial^\mu \bar K^0)\right\},
\end{eqnarray}
\begin{eqnarray}
   S_{mass}&\!\!\! =\!\!\!& -\int d^4x \left\{
   M^2_{\sigma}(m_u,m_u)(\sigma_u^2 +\sigma_d^2 +\sigma^+\sigma^-)+
   M^2_{\sigma}(m_s,m_s)\sigma_s^2 \right.
   \nonumber\\
   &\!\!\! +\!\!\!& M^2_{\sigma}(m_u,m_s)(\delta^0\bar\delta^0 
   +\delta^+\delta^-)
   \nonumber\\
   &\!\!\! +\!\!\!& M^2_{\pi}(m_u,m_u)(\phi_u^2 +\phi_d^2 +\pi^+\pi^-)
   +M^2_{\pi}(m_s,m_s)\phi_s^2 
   \nonumber\\
   &\!\!\! +\!\!\!& \left. M^2_{\pi}(m_u,m_s)(K^0\bar K^0 +K^+K^-)
   \right\}.
\end{eqnarray}
If we use the gap equations in order to eliminate the quadratic divergencies in the expressions for the renormalized masses
$\mu^2_{\sigma,\pi}(m_i,m_j)$, Eq. (\ref{Ms}), it is possible to define the renormalized coupling $C_i$ through
\begin{eqnarray}
\label{cur}
C_{u}&=&\frac{c_{u0}}{g_0^2}=m_u\frac{\mu^2_\pi}{g_\pi^2}
\nonumber\\
C_{s}&=&\frac{c_{s0}}{g_0^2}
\nonumber\\
&=&m_s \left(\frac{2\mu_{\kappa}^2}{g_{\kappa}^2}-\frac{\mu_\pi^2}{g_\pi^2}\right.
\nonumber\\
&-&\left. 2(m_u-m_s)^2(\frac{\lambda_q}{g_\pi^2}-4N_cF(m_u,m_s,0;\zeta^2))\right).
\end{eqnarray}
Using (\ref{ukup}) one obtains also
\begin{equation}
\label{cups}
C_{u}+C_{s}=(m_u+m_s)\frac{\mu_k^2}{g_{\kappa}^2}.
\end{equation}
Now we write out the renormalized propagators for pions and kaons, since they will be used in the remaining,
\begin{eqnarray}
\label{propa}
\Delta_\pi^{-1}(p^2)&=&p^2-\mu_\pi^2+4N_cg_\pi^2 F_{fin}(m_u,m_u,p^2;\zeta^2)\\
\Delta_{\kappa}^{-1}(p^2)&=&p^2-\mu_{\kappa}^2+ 4N_cg_{\kappa}^2\Sigma_{fin}(m_u,m_s,p^2;\zeta^2)
\end{eqnarray}
with the finite momentum dependent contributions
\begin{eqnarray}
F_{fin}(m_u,m_u,p^2;\zeta^2)&=-p^2 (F(m_u,m_u,p^2;\zeta^2)-F(m_u,m_u,0;\zeta^2))
\end{eqnarray}
where in particular $F(m_i,m_i,0;m_i)=0$,
\begin{eqnarray}
   \Sigma_{fin}(m_u,m_s,p^2;\zeta^2) &\!\!\! =\!\!\!&
   [(m_s-m_u)^2-p^2]\{F(m_u,m_s,p^2;\zeta^2)-F(m_u,m_s,0;\zeta^2)\}
   \nonumber\\
   &\!\!\! -\!\!\!& p^2 (m_s-m_u)^2 F'(m_u,m_s,0;\zeta^2)
\end{eqnarray}
where use has been made of the normalization conditions
\begin{equation}
\label{nor}
   \Delta_{\pi, {\kappa}}^{-1}(0)= -\mu_{\pi, {\kappa}}^2,
   \qquad
   \frac{d\Delta_{\pi, {\kappa}}^{-1}(p^2)}{d p^2} |_{p^2=0} = 1.
\end{equation}
We obtain finally the physical pseudoscalar masses as zeros of these propagators
\begin{eqnarray}
\label{mphys}
m_\pi^2&=&\mu_\pi^2-4N_cg_\pi^2F_{fin}(m_u,m_u,m_\pi^2;\zeta^2)
\nonumber\\
m_{\kappa}^2&=&\mu_{\kappa}^2- 4N_cg_{\kappa}^2\Sigma_{fin}(m_u,m_s,m_{\kappa}^2;\zeta^2)
\end{eqnarray}

At the physical meson masses we obtain the following pseudoscalar quark couplings
\begin{eqnarray}
\label{gpqq}
g_{\pi qq}^{-2}&=&g_{\pi}^{-2}+4N_c\frac{d F_{fin}(m_u,m_s,p^2;\zeta^2)}{dp^2}|_{p^2=m_\pi^2}
\nonumber\\
g_{\kappa qq}^{-2}&=&g_{\kappa}^{-2}+4N_c\frac{d\Sigma_{fin}(m_u,m_s,p^2;\zeta^2)]}{dp^2}|_{p^2=m_\kappa^2}
\end{eqnarray}

{\bf 3.A - The $\beta$ -function}
\vspace{0.5 cm}

In order to illustrate the role of the arbitrary scale $\zeta^2$ introduced in the previous sections, we evaluate the $\beta$-function
for the pseudoscalars
\begin{equation}
\label{beta}
\beta_\pi=\zeta\frac{\partial}{\partial \zeta} g_\pi(\zeta^2).
\end{equation}
Using the Eq. (\ref{gp}) we get to one loop order
\begin{equation}
\label{bepi}
\beta_\pi=\frac{N_c g_\pi^3}{4 \pi^2}.
\end{equation}
>From the renormalization group equation (\ref{beta}) and (\ref{bepi}) one solves for $g_\pi$ and gets
\begin{equation}
\label{gepi}
g_\pi^2(\zeta'^2)=\frac{g_\pi^2(\zeta^2)}{1- \frac{N_c}{4 \pi^2} g_\pi^2(\zeta^2){\it ln}(\frac{\zeta'^2}{\zeta^2})}.
\end{equation}
Analogously to QED at one loop level, a Landau pole appears in equation (\ref{gepi}) and it corresponds to a pole in the large
Euclidean momentum region of the pion propagator. This is easily seen by  expanding for large negative $p^2$ the inverse of the
renormalized pion propagator, Eq. (\ref{propa})
\begin{equation}
\Delta_\pi^{-1}\sim |p^2|(\frac{N_c}{4\pi^2}g_\pi^2 {\it ln}\frac{|p^2|}{\zeta^2}-1)+...
\end{equation}
\vspace{1cm}

{\bf 4. Coupling to external fields}
\vspace{.5cm}

In order to make contact to phenomenology an important issue is the calculation of electromagnetic form factors and weak
decay constants. We treat the vector and axial currents as external classical fields. We extract these quantities from a
generalized expression for the Dirac operator, now containing the vector $V_\mu$ and axial-vector $A_\mu$ fields.
The previous expression for $D^{\dagger}D$ is now replaced by

\begin{eqnarray}
D_E^{\dagger}D_E&=&m^2-d_\mu^2 +i g_0\gamma_\mu(\Delta_\mu\sigma_0-i\gamma_5\Delta_\mu\pi_0)-
\frac{i}{4}[\gamma_\mu,\gamma_\nu]\Gamma_{\mu\nu}
\nonumber\\
         &+&g_0^2((\sigma^2_0+\{\sigma_0,m/g_0\}+\pi_0^2+i\gamma_5[\sigma_0+m/g_0,\pi_0]),
\end{eqnarray}

where
\begin{equation}
-d_\mu^2=-\partial_\mu^2 +2i\Gamma_\mu\partial_\mu+i\partial_\mu \Gamma_\mu +\Gamma_\mu^2
\end{equation}
and $\Gamma_\mu=V_\mu+\gamma_5 A_\mu$, $\Gamma_{\mu\nu}=V_{\mu\nu}+\gamma_5 A_{\mu\nu}$.
The covariant derivatives are
\begin{eqnarray}
\label{covar}
   \nabla_\mu\sigma_0&\!\!\! =\!\!\! &\partial_\mu 
   \sigma_0+\{A_\mu,\pi_0\}-i[V_\mu,\sigma_0+m/g_0] \\
   \nabla_\mu\pi_0&\!\!\! =\!\!\!&\partial_\mu \pi_0
   -\{A_\mu,\sigma_0+m/g_0\}-i[V_\mu,\pi_0] \\
   V_{\mu\nu}&\!\!\! =\!\!\!&\partial_\mu V_\nu-\partial_\nu V_\mu 
   - i[A_\mu,A_\nu]-i[V_\mu,V_\nu] \\
   A_{\mu\nu}&\!\!\! =\!\!\!&\partial_\mu A_\nu-\partial_\nu A_\mu 
   - i[V_\mu,A_\nu]-i[A_\mu,V_\nu].
\end{eqnarray}

The calculation of $f_\pi$ and $f_\kappa$ to one loop order is now straightforward, after collecting the relevant vertices involving
one pseudoscalar and one axial-vector fields. Since we are treating the external currents classically, we disregard all quantum
 dynamical corrections to the vector fields. Also the kinetic terms proportional to $f_0^2$ involving the scalar and pseudoscalar
 fields in Eq.
(\ref{action}) have to be minimally coupled to the external currents through the covariant derivatives Eq. (\ref{covar}).
We obtain the bare $f_\pi^0$ as
\begin{eqnarray}
\label{fpi}
f^0_\pi(m_\pi^2)&=&\frac{m_u}{g_0}(-4N_cg_0^2I_{log}(\zeta^2)+f_0^2-4N_cg_0^2F(m_u,m_u,m_\pi^2,\zeta^2))
\nonumber\\
&=&\frac{m_u}{g_0}Z_\pi^{-1}(m_u,m_u)\left(1-4N_cg_\pi^2(\alpha-\frac{F_{fin}(m_u,m_u,m_\pi^2,\zeta^2)}{m_\pi^2}\right)
\end{eqnarray}

In order to obtain the Goldberger-Treiman relation of chiral symmetry it is
necessary to avoid the ambiguous $\alpha$ term present in Eq.(\ref{fpi}).
This can be done by either choosing $\alpha =0$ at this point \cite{Dias2006}
or by introducing the following counter-terms
\begin{eqnarray}
\label{Lct}
L_{c.t.}&=&\frac{f^{2}g_0^2}{4}\mbox{Tr}(- \partial _{\mu }\pi \{A_{\mu },\sigma_0 +\frac{m}{g_0}\}-i\partial
_{\mu }\pi_0 [V_{\mu },\pi ]  \label{ctinv} \\
&&+\partial _{\mu }\sigma_0 \{A_{\mu },\pi_0 \}-i\partial _{\mu }\sigma_0 [V_{\mu
},\sigma_0 +\frac{m}{g_0}]),  \nonumber
\end{eqnarray}
with a parameter $f^2$ to be fixed appropriately.
This is the minimal chiral and gauge invariant combination available which contains a term that can absorb the $\alpha$ ambiguity. For the present purpose this is enough, although one might need to consider higher than cubic field interaction terms as well to realize the complete renormalization program. Note that this counterterm has consequences for the scalar-vector (SV) sector, which will be discussed below.
Including the terms from (\ref{ctinv}) that contribute to the pion weak decay
constant (proportional to $A_{\mu }\partial ^{\mu }\pi $), we obtain
\begin{equation}
f_{\pi }^{0}(m_{\pi }^{2})=\frac{m_{u}}{g_{0}}Z_{\pi
}^{-1}(m_{u},m_{u})\left(1+f^{2}g_\pi^2 -4N_{c}g_{\pi }^{2}(\alpha-\frac{F_{fin}(m_{u},m_{u},m_{\pi
}^{2},\zeta ^{2}}{m_\pi^2})\right). \label{fpi2}
\end{equation}

Using that
\begin{equation}
-\frac{4N_cg_\pi^2}{m_\pi^2}F_{fin}(m_{u},m_{u},m_{\pi}^{2},\zeta ^{2})=1-\frac{\mu_\pi^2}{m_\pi^2}
\end{equation}
and the relation between the physical and bare pion field at the physical pion mass, $\pi=\pi^{ph}\frac{g_{\pi qq}}{g_0}$,
we get the renormalized weak pion decay constant
\begin{equation}
\label{fpifin}
f_\pi(m_\pi^2)=m_u\frac{g_{\pi qq}}{g_\pi^2}\frac{\mu_\pi^2}{m_\pi^2}
\end{equation}
provided
\begin{equation}
\label{ff}
f^{2}=4N_c \alpha
\end{equation}
in Eq. (\ref{fpi2}), as a consequence of requiring chiral symmetry.

Now, for the kaon weak decay constant one obtains

\begin{eqnarray}
\label{fk}
f_{\kappa}^{0}(m_{\kappa}^{2}) &=&2N_{c}\frac{g_{0}}{m_{\kappa}^{2}}
(m_{s}-m_{u})\left[(I_{q}(m_{u}^2)-I_{q}(m_{s}^2)+m_{\kappa}^{2} \alpha) \right.  \\
&+&\left. (m_{s}^{2}-m_{u}^{2})(I_{log}(\zeta ^{2})+F(m_{u},m_{s},m_{\kappa}^{2};\zeta
^{2})\right]  \nonumber \\
&+&\frac{(m_{u}+m_{s})}{2g_{0}}\left[-4N_{c}g_{0}^{2}I_{log}(\zeta ^{2})+f_{0}^{2}+f^2 g_0^2 \right.
\nonumber \\
&-& \left. 4N_{c}g_{0}^{2}F(m_{u},m_{s},m_{\kappa}^{2};\zeta ^{2})\right],  \nonumber
\end{eqnarray}
The term proportional to $(m_u+m_s)$ is the term which in the limit of equal masses reproduces $f_\pi$.
We see that the introduction of the counter-terms (\ref{ctinv}) in the model Lagrangian, with $f^{2}=4N_c \alpha$
allows the $(m_{u}+m_{s})$ part of (\ref{fk}) to be renormalized in the same way
we did for $f_{\pi }$. However there remains an $\alpha$ dependent piece, proportional to the difference of the quark masses, which cannot be removed by the counterterm, nor by renormalization.
It is interesting to note that the ambiguity $\alpha$ can neither be fixed by the Ward identity
related to the pseudoscalar/axialvector radiative amplitude
\begin{equation}
\label{IW}
   p^{\mu }\Pi_{\mu }^{PA}=(m_{u}+m_{s})\Pi^{PP}+4m_uT(p^2,m_u)+4m_sI_q(m_s^2),
\end{equation}
\noindent where
\begin{equation}
   {\Pi }_{\mu }^{PA} = i\int_{\Lambda }\frac{d^{4}k}{(2\pi )^{4}}
   \mbox{Tr}\left(\gamma_{5}\frac{1}{\not{k}-m_{s}}\gamma_{\mu }
   \gamma_{5}\frac{1}{\not{k}\ -\not {p}-m_{u}}\right)
\end{equation}
\noindent and $T(p^{2},m_{i})$ given in Eq. (\ref{arb}).
One can verify that Eq.(\ref{IW}) is fulfilled without fixing $\alpha$,
as long as the ambiguity present in the pseudoscalar/axialvector amplitude and the
one present in the pseudoscalar/pseudoscalar amplitude are the same.
One concludes that in the PA sector alone an ambiguity remains undetermined, which is compatible with the Goldberger-Treiman relation and modifies the PCAC relation only as a next to leading order correction in the symmetry breaking term and is leading correction in flavor symmetry breaking.
This conclusion is however fallacious, as the chiral symmetry partner sector SV has to be analyzed in parallel.
By fixing $\alpha$ through Eq. (\ref{ff}),
Eq.(\ref{ctinv}) introduces counterterms in the
scalar-vector sector
\begin{equation}
   L_{ctSV}=(m_{s}-m_{u})f^{2}\left(
   V_{\mu }^{+}\partial_{\mu }\delta^{-}
  -V_{\mu }^{-}\partial_{\mu }\delta^{+}
  +V_{\mu }^{^{\prime }+}\partial_{\mu }\overline{\delta }_{0}
  -V_{\mu }^{^{\prime }-}\partial_{\mu }\delta _{0}\right),  
\label{Lctsv}
\end{equation}
where $V_{\mu }^{+}=V_{4\mu }-iV_{5\mu }$, $V_{\mu }^{-}=V_{4\mu }+iV_{5\mu }
$, $V_{\mu }^{^{\prime }+}=V_{6\mu }-iV_{7\mu }$ and $V_{\mu }^{^{\prime
}-}=V_{6\mu }+iV_{7\mu }$.  These counterterms absorb the ambiguous $\alpha$ proportional terms which are generated radiatively through
\begin{equation}
   \Pi _{SV}^{\mu ,ab}=\frac{i}{4}\int\frac{d^{4}k}{(2\pi )^{4}}
   \mbox{Tr} \left( \frac{1}{\not {k}-m}
   \gamma^{\mu }\lambda _{a}
   \frac{1}{\not {k}\ -\not {p}-m}
   \lambda _{b}\right).  
\label{PiSVab}
\end{equation}

However, opposite to the PA case, these counterterms are proportional to the difference of the quark masses and will leave untouched an ambiguous term proportional to the sum of quark masses. This latter appears in a similar fashion as in Eq.(\ref{fk}) through the difference of quadratic divergencies with different external momentum dependence. This term violates chiral symmetry, since the SV amplitude must vanish in the limit of equal quark masses.
We conclude therefore that the ambiguity $\alpha$ must be set to zero in this sector. This immediately implies that it must be zero as well in the PA sector, as the counterterms in the two sectors are linked through chiral symmetry.

We consider from now on the case with $\alpha=0$.

Returning to Eq.(\ref{fk}), we observe that the quadratic divergencies can be removed through the gap equations and the
other divergencies by renormalization.
Using Eqs. (\ref{gkgp}-\ref{fmums0}) and (\ref{mphys}) $f_{\kappa}$ reduces to
\begin{eqnarray}
\label{fkfin}
f_{\kappa}(m_{\kappa}^2)&=&g_{{\kappa}qq}\frac{(m_u+m_s)}{2} \frac{\mu_{\kappa}^2}{m_{\kappa}^2 g_{\kappa}^2}
\end{eqnarray}
At this stage it is useful to see how the expressions for $f_\pi$ and $f_{\kappa}$ comply with PCAC. For that
we use the explicit symmetry breaking term of the Lagrangian in Eq. (\ref{action}), \cite{Gasiorowicz:1969}
\begin{equation}
\label{symbr}
\delta {\cal L}_{SB}=\frac{1}{2 g_0}\delta Tr(c_0\sigma_0)
\end{equation}
to obtain for instance for one isotopic component of the pion field
\begin{equation}
\label{pionbr}
\frac{\partial \delta  {\cal L}_{SB}}{\partial \beta_1}=-2\pi_1\frac{C_{0u}}{g_0}=-2\pi_1^{ph}g_{\pi qq} C_u
\end{equation}
and for one flavour component of the kaon
\begin{equation}
\label{kaonbr}
\frac{\partial \delta  {\cal L}_{SB}}{\partial \beta_4}=-K_4\frac{C_{0u}+C_{0s}}{g_0}=-K_4^{ph} g_{Kqq} (C_u+C_s)
\end{equation}
Upon using here Eqs. (\ref{cur}) and (\ref{cups}) and the expressions (\ref{fpifin}) and (\ref{fkfin}) one obtains
\begin{equation}
\label{spi}
\frac{\partial \delta  {\cal L}_{SB}}{\partial \beta_1}=-2\pi_1^{ph} m_\pi^2 f_\pi(m_\pi^2)
\end{equation}
\begin{equation}
\label{ska}
\frac{\partial \delta  {\cal L}_{SB}}{\partial \beta_4}=-2K_4^{ph} m_{\kappa}^2 f_{\kappa}(m_{\kappa}^2).
\end{equation}
So, the pion and the kaon fulfill exactly PCAC.
\vspace{1cm}

{\bf{5. Numerical Results}}

\vspace{0.5cm}
In  this section, we fit the model parameters to reproduce the pion and
kaon masses, as well as the pion weak decay constant, in order to obtain a
numerical estimative to the kaon weak decay constant and the kaon
electromagnetic form factor. The model parameters are fixed by means of
Eq. (\ref{mphys}) and (\ref{fpi}) in order to reproduce $m_{\pi }=139MeV$,
$m_{\kappa }=494MeV$ and $f_{\pi }=93.3MeV$, with the renormalization
constants $g_{\kappa }^{2}$ and $\mu _{\kappa }^{2}$ given by
(\ref{gkgp}) and (\ref{ukup}).

As pointed in ref. \cite{andre:1999}, the action defined by
(\ref{Sterms}) gives the same formal results as the conventional NJL model (with
a $\Lambda \rightarrow \infty $ cut-off) in the case where $\lambda _{q}=1$. In
particular, the Nambu relation in the chiral limit $m_{\sigma }=2m$
holds only in the $\lambda _{q}=1$ case. We will proceed the parameters
fitting with $\lambda _{q}=1$, and will return to this point latter. We also
use $ m_{u}=210MeV$ and $m_{u}=350MeV$ , values largely employed on
literature, generating the two set of parameters showed on table I.

\begin{table}[tbp]

\begin{center}

\begin{tabular}{|c|c|c|}

\hline

&  &  \\

Set & $m_{u}=210MeV$ & $m_{u}=350MeV$ \\

&  &  \\ \hline

&  &  \\

$g_{\pi }$ & $2.250$ & $3.752$ \\

$\mu _{\pi }$ & $141.1MeV$ & $141.0MeV$  \\

$m_{s}$ & $479.0MeV$ & $601.7MeV$ \\

$f_{\kappa }$ & $114.0MeV$ & $71.6MeV$ \\

$<r_{\kappa}^2>^{1/2}$ & $0.573fm$ & $0.608fm$\\

&  & \\ \hline

\end{tabular}

\end{center}

\caption{Parameters of the model, fitted to reproduce $m_{\pi}=139MeV$,
$m_{\kappa}=494MeV$ and $f_{\pi}=93.3MeV$, with $m_{u}=210MeV$ and $m_{u}=350MeV$.}

\end{table}

With these choices for the parameters, and with $N_{c}=3$, we obtain for the kaon weak decay constant, $f_{\kappa}=114.2MeV$
(for $m_{u}=210MeV$) and $f_{\kappa }=71.6MeV$ (for $m_{u}=350MeV$).

The value of $f_{\kappa }$ estimated by the set of parameters
corresponding to $m_{u}=210MeV$ is only $\sim 3\%$ lower than the experimental
results. Nevertheless, for $m_{u}=350MeV$ the kaon weak decay constant is still
away from its experimental value, as already known.

We also evaluate the pion and kaon electromagnetic form factors, and
the corresponding pion and kaon charge radius in the space-like region,
given by
\begin{eqnarray}
\label{ffk}
F_{\kappa } &=&-N_{c}g_{\kappa qq}^{2}[\frac{1}{g_{\pi }^{2}}
+F(m_{u}^{2},m_{u}^{2},q^{2};\zeta ^{2})-(m_{\kappa
}^{2}-(m_{s}-m_{u})^{2})F^{\prime }(m_{s},m_{u},m_{\kappa }^{2};\zeta
^{2})+
\nonumber \\
&&I_{t}(q^{2},m_{s},m_{u})]-N_{c}g_{\kappa qq}^{2}[\frac{1}{g_{\pi
}^{2}}
+F(m_{s}^{2},m_{s}^{2},q^{2};\zeta ^{2})-  \nonumber \\
&&(m_{\kappa }^{2}-(m_{s}-m_{u})^{2})F^{\prime }(m_{u},m_{s},m_{\kappa
}^{2};\zeta ^{2})+I_{t}(q^{2},m_{u},m_{s})],
\end{eqnarray}
and

\begin{equation} \label{rk}
<r_{\kappa }^{2}>= -6\frac{dF_{\kappa }(Q^{2})}{dQ^{2}}\arrowvert_{Q^{2} \rightarrow 0}
\end{equation}
where
\begin{equation}
I_{t}(q^{2},m_{u},m_{s})=i\int \frac{d^{4}k}{(2\pi )^{4}}\frac{1}{
(k^{2}-m_{u}^{2})[(k-k_{1})^{2}-m_{s}^{2}][(k-k_{2})^{2}-m_{s}^{2}]}
\end{equation}
with

\begin{equation}
k_{1}^{2}=k_{2}^{2}=m_{\kappa }^{2}
\end{equation}
and

\begin{equation}
k_{1}.k_{2}=m_{\kappa }^{2}-\frac{q^{2}}{2}.
\end{equation}
Also, in Eq.(\ref{rk}) and in figures 1-3, $Q^2=-q^2$ as usual.

For the pion electromagnetic form factor and pion radius, the
expressions are given by (\ref{ffk}) and (\ref{rk}) in the $m_{s}=m_{u}$ limit.

\begin{figure}

\centerline{\psfig{file=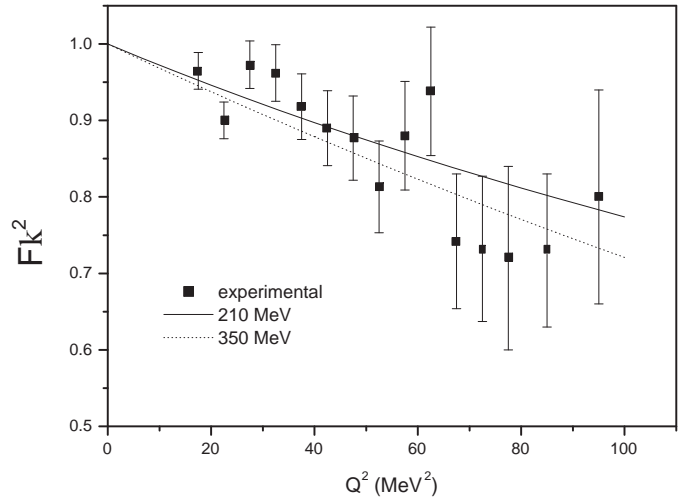,width=10cm}}

\vspace*{8pt}

\caption{ Kaon electromagnetic form factor
in the space-like region compared with experimental data \cite
{Amendolia86} . Solid line corresponds to the set of
parameters fitted with $m_{u}=350MeV$, dashed line corresponds to
$m_{u}=210MeV$. }

\end{figure}

\begin{figure}

\centerline{\psfig{file=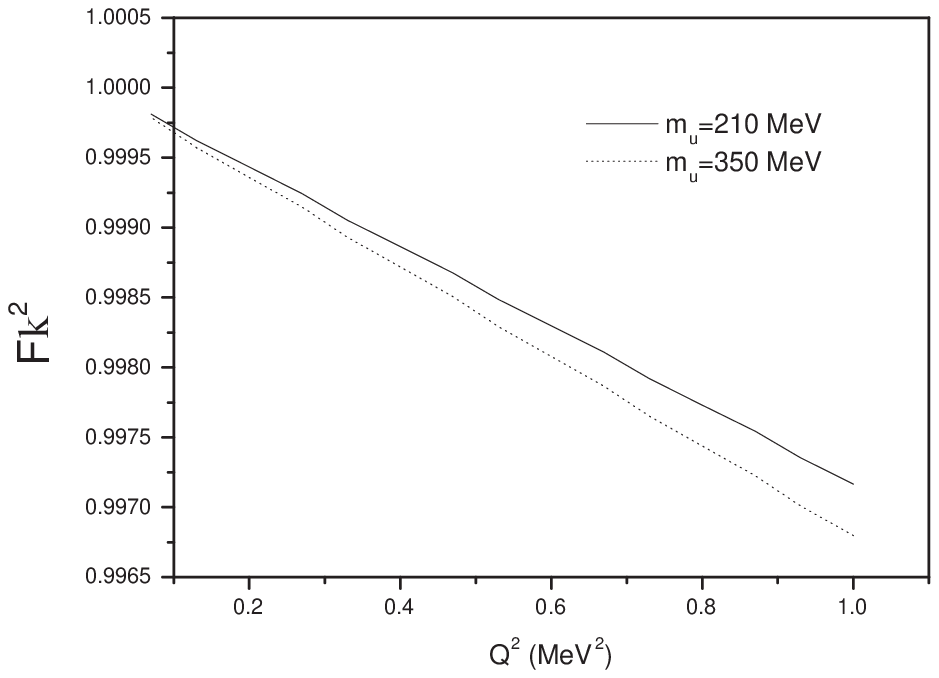,width=10cm}}

\vspace*{8pt}

\caption{ Kaon electromagnetic form factor
in the space-like region. Solid line corresponds to the set of
parameters fitted with $m_{u}=350MeV$, dashed line corresponds to
$m_{u}=210MeV$. The figure shows up the difference in the slope of the curves. }

\end{figure}

In figure 1 we plot the kaon electromagnetic form factor in the space-like region, evaluated in
the present model compared with experimental data. This result shows a good
agreement between the model and experiment. In figure 2 we focused the
region near $Q^{2}=0$, that highlights the differences in the slopes of
the $m_{u}=210MeV$ and $m_{u}=350MeV$ fittings, and explains why the
difference in the kaon charge radius evaluated with these two sets of parameters
is greater than it seems from fig. 1. Figure 3 presents the Pion
electromagnetic form factor.

\begin{figure}

\centerline{\psfig{file=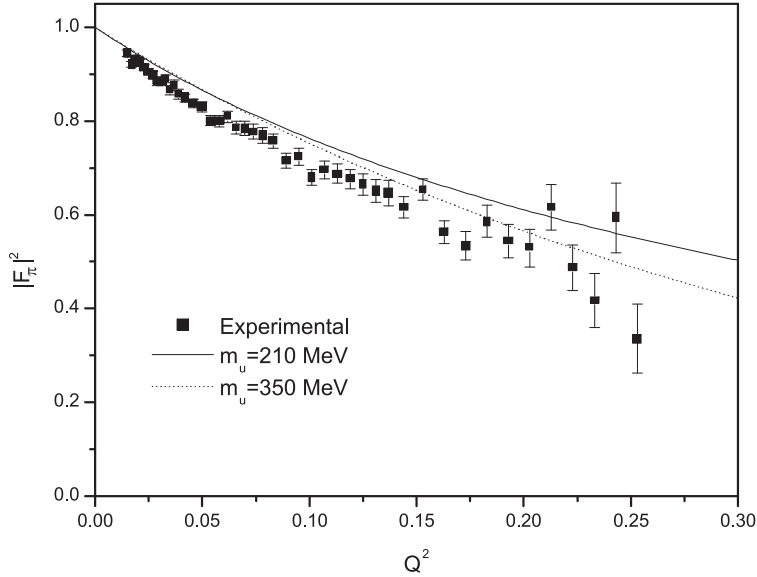,width=10cm}}

\vspace*{8pt}

\caption{ Pion electromagnetic form
factor in the space-like region  compared with experimental data \cite{Amendolia84}.
Solid line corresponds to the set of
parameters fitted with $m_{u}=350MeV$, dashed line corresponds to $
m_{u}=210MeV$. }

\end{figure}

\begin{figure}

\centerline{\psfig{file=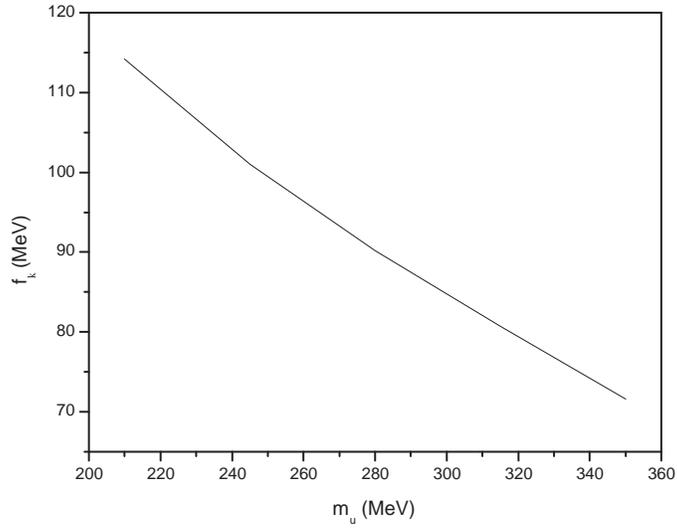,width=10cm}}

\vspace*{8pt}

\caption{ Kaon weak decay constant for different quark up constituent
masses.  }

\end{figure}

In figure 4 we show the Kaon weak decay constant for the up
constituent mass in the range $210MeV<m_{u}<350MeV$. We can see that $f_{\kappa}$
decreases smoothly as $m_{u}$ increases.

In order to investigate if a different choice of $\lambda _{q}$ could
provide a better fitting of the kaon weak decay constant in the $m_{u}=350MeV$ case, we also plot, in figure 5,
the dependence of the kaon weak decay constant with $\lambda _{q}$. We verify that it is still not
possible to fit the Kaon weak decay constant with its experimental value for any value of $\lambda_q$.
In fact, it is not possible to adjust the parameters in order to reproduce $m_{\kappa }=494MeV$ for
$\lambda _{q}<0.93$. For $0.93<\lambda _{q}<1.00$ we have $f_{\kappa }<71.6MeV$, and as $\lambda _{q}$ tends to infinity,
the kaon weak decay constant tends to $f_{\kappa}=94.6MeV$ , and the constituent strange mass, which also depends
on $\lambda _{q}$, runs from $m_{s}=644MeV$ to $m_{s}=350MeV$.

\begin{figure}

\centerline{\psfig{file=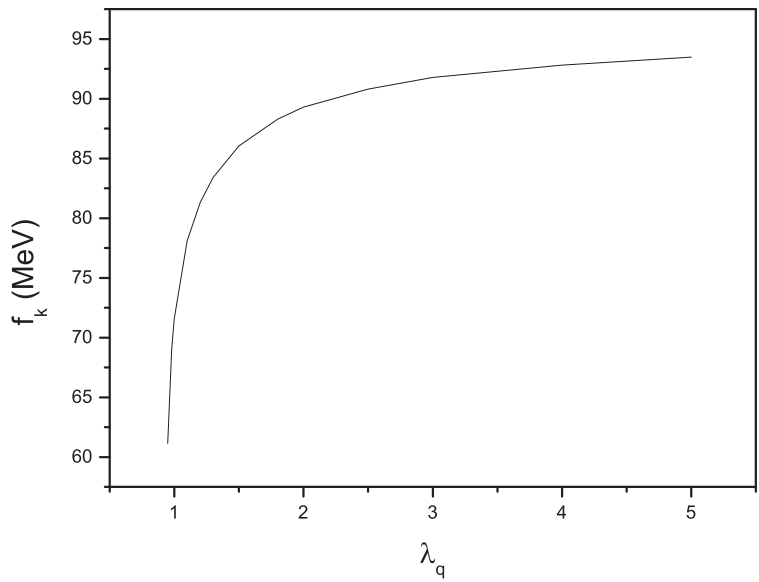,width=10cm}}

\vspace*{8pt}

\caption{ Kaon weak decay constant as a function of $\lambda _{q}$ for
$m_{u}=350MeV$. }

\end{figure}

\vspace{1cm}


{\bf 6. Conclusions}
\vspace{.5cm}

We investigated the role of ambiguities associated with finite surface terms of the
type of differences between logarithmic divergencies in the LSM with fermions. We found in
the pseudoscalar-axialvector sector an arbitrary c-number which cannot be absorbed by renormalization
nor fixed through the Goldberger-Treiman relation or PCAC.
In this process, however, necessary chiral and gauge invariant counterterms were introduced which link the pseudoscalar-axialvector sector with the scalar-vector one.  The counterterms have a common coefficient which is fixed by the value assigned to the arbitrary c-number. If this number is chosen to be non-zero, it induces a symmetry breaking term in the scalar-vector sector. We must conclude therefore that the apriori arbitrary c-numbers are finally removed in order not to violate chiral symmetry in this sector.
Further, we have obtained numerical results which show a good agreement between the model and experimental data
for the pion and kaon electromagnetic form factors, even for higher values of the quark up constituent mass.
The kaon weak decay constant for $m_{u}=350MeV$, nevertheless, cannot be adjusted close to its
experimental value in the range of the other free model parameters.

\vspace{.5cm}

{\bf Acknowledgements}
\vspace{.5cm}

This work has been supported by grants provided by Funda\c c\~ao para
a Ci\^encia e a Tecnologia, POCTI/35304/FIS/2000, POCTI/FNU/50336/2003, POCI/FP/63412/2005 and PRAXIS XXI/BCC/4301/94.
This research is part of the EU integrated infrastructure initiative
HadronPhysics project under contract No.RII3-CT-2004-506078.
A.A. Osipov also gratefully acknowledges the
Funda\ca o Calouste Gulbenkian for financial support and A.L. Mota acknowledges financial support
from CNPq/Brazil and CAPES/Brazil.
\vspace{0.5cm}

{\bf Appendix}
\vspace{.5cm}

Consider first the following example. The surface-term integral
\begin{equation}
\label{Y}
   Y_{\mu\nu\alpha\beta}(m^2)=i\int \frac{d^4p}{(2\pi)^4}
   \left[\frac{g_{\mu\nu\alpha\beta}}{(p^2-m^2)^2}
        -24\frac{p_\mu p_\nu p_\alpha p_\beta}{(p^2-m^2)^4}\right]
\end{equation}
does not depend on $m^2$, i.e.
\begin{equation}
\label{0}
  \frac{dY_{\mu\nu\alpha\beta}}{dm^2}=2i\int \frac{d^4p}{(2\pi)^4}
   \left[\frac{g_{\mu\nu\alpha\beta}}{(p^2-m^2)^3}
        -48\frac{p_\mu p_\nu p_\alpha p_\beta}{(p^2-m^2)^5}\right]=0,
\end{equation}
where we use the standard notation
$$
   g_{\mu\nu\alpha\beta}=g_{\mu\nu}g_{\alpha\beta}
   +g_{\mu\alpha}g_{\nu\beta}+g_{\mu\beta}g_{\nu\alpha}\ .
$$

Let us generalize Eq.(\ref{0}). For that, note that
in the space of $d=2\omega$ dimensions one has
$$
   g_{\mu_{1}\mu_{2}\ldots \mu_{2n}}g^{\mu_1\mu_2}\ldots
   g^{\mu_{2n-1}\mu_{2n}}
   =2^n\frac{\Gamma (\omega +n)}{\Gamma (\omega )}.
$$
Therefore one obtains the identity
\begin{eqnarray}
\label{eq77}
   \int d^{2\omega}p
   \frac{p_{\mu_{1}} p_{\mu_{2}}\ldots p_{\mu_{2n}}
   }{(p^2-m^2)^A}
   &=&\frac{i\pi^\omega \Gamma (A-\omega-n)}{(m^2)^{A-\omega-n}
   2^n\Gamma (A)}
   g_{\mu_{1}\mu_{2}\ldots \mu_{2n}}\\ \nonumber
   &=&\frac{\Gamma (A-n)}{2^n \Gamma (A)}
    \int d^{2\omega}p \frac{g_{\mu_{1}\mu_{2}\ldots \mu_{2n}}
    }{(p^2-m^2)^{A-n}}
\end{eqnarray}
Eq.( \ref{eq77}) can be rewritten as
\begin{equation}
   \int d^{2\omega}p\left[
   \frac{g_{\mu_{1}\mu_{2}\ldots \mu_{2n}}
    }{(p^2-m^2)^{A-n}}-
   \frac{2^n\Gamma (A)}{\Gamma (A-n)}
   \frac{p_{\mu_{1}} p_{\mu_{2}}\ldots p_{\mu_{2n}}
   }{(p^2-m^2)^A}\right]=0.
\end{equation}
Integrating now with respect to $m^2$ we obtain the most general form
of all possible surface terms
\begin{equation}
   S^{(A-2)}_{\mu_{1}\mu_{2}\ldots \mu_{2n}}=i
   \int \frac{d^{2\omega}p}{(2\pi )^4}
   \left[
   \frac{g_{\mu_{1}\mu_{2}\ldots \mu_{2n}}
    }{(p^2-m^2)^{A-n-1}}-
   \frac{2^n\Gamma (A-1)}{\Gamma (A-n-1)}
   \frac{p_{\mu_{1}} p_{\mu_{2}}\ldots p_{\mu_{2n}}
   }{(p^2-m^2)^{A-1}}\right]
\end{equation}


\begin{thebibliography}{99}
\bibitem{Jackiw:2000} R. Jackiw, Int. J. Mod. Phys. B {\bf 14}, 2001 (2000), hep-th/9903044.
\bibitem{Bell:1969}J.S. Bell and R. Jackiw, Nuovo Cimento {\bf 60A}, 47 (1969).
\bibitem{Adler:1969} S.L. Adler, Phys. Rev. {\bf 177}, 2426 (1969).
\bibitem{anom} A.P. Ba\^eta Scarpelli, M. Sampaio, M.C. Nemes, and B. Hiller, Phys. Rev. D {\bf 64},
        046013 (2001), hep-th/0102108.
\bibitem{orimar:1999} O.A. Battistel, PhD thesis, Federal University of Minas Gerais, Brazil.
\bibitem{baeta1:2001} A.P. Ba\^eta Scarpelli, M. Sampaio, and M.C. Nemes, Phys. Rev. D {\bf 63}, 046004
        (2001), hep-th/0010285;
        M. Sampaio,A.P. Ba\^eta Scarpelli, B. Hiller, A. Brizola, M.C. Nemes, and S. Gobira,
        Phys. Rev. D {\bf 65}, 125023 (2002), hep-th/0203261.
\bibitem{gobira:2004} S.R. Gobira and M.C. Nemes, Int. J. Theor. Phys. {\bf 42}, 2765 (2003).
\bibitem{carneiro:2004} D. Carneiro, A.P. Ba\^eta Scarpelli, M. Sampaio, M.C.
Nemes, JHEP 0312 (2003) 044, hep-th/0309188.
\bibitem{souza:2005}Leonardo A.M. Souza, Marcos Sampaio, M.C. Nemes, Phys. Lett. B {\bf 632},717 (2006), hep-th/0510017.
\bibitem{Levy:1960} M. Gell-Mann and M. L\'evy, Nuov. Cim. {\bf 16}, 705 (1960).
\bibitem{Bonneau2000} G. Bonneau, Nucl. Phys. B {\bf 593}, 398 (2001), hep-th/0008210.
\bibitem{Hooft:1976} A. M. Polyakov, Phys.
Lett. B {\bf 59}, 82 (1975); {\it idem}        Nucl. Phys. {\bf B120}, 429 (1977).
        A. A. Belavin, A. M. Polyakov, A. Schwartz and Y. Tyupkin,
        Phys. Lett. B {\bf 59}, 85 (1975);
        G. 't Hooft, Phys. Rev. Lett. {\bf 37}, 8 (1976); Phys. Rev. D
        {\bf 14}, 3432 (1976);
        C. Callan, R. Dashen and D. J. Gross, Phys. Lett. B {\bf 63},
        334  (1976);
        R. Jackiw and C. Rebbi, Phys. Rev. Lett. {\bf 37}, 172 (1976);
        S. Coleman, {\it The uses of instantons} (Erice Lectures, 1977);
        G. 't Hooft, hep-th/9903189.
\bibitem{Gasiorowicz:1969} S.Gasiorowicz, D.A. Geffen, Rev. Mod. Phys. {\bf 41}, 531 (1969).
\bibitem{Nambu:1961} Y.Nambu, G. Jona-Lasinio, Phys. Rev. {\bf 122}, 345 (1961);
                     Y.Nambu, G. Jona-Lasinio, Phys. Rev. {\bf 124}, 246 (1961);
                     V.G. Vaks, A.I. Larkin, Zh. Eksp. Teor. Fiz. {\bf 40}, 282 (1961).
\bibitem{Schwinger:1951} J. Schwinger, Phys. Rev. {\bf 82},664 (1951).
\bibitem{DeWitt:1965} B. DeWitt, Dynamical Theory of Groups and Fields, Gordon and Breach, N.Y.,1965.
\bibitem{Ball:1989} R.D. Ball, Phys. Rep. {\bf 182}, 1 (1989) (and references therein).
\bibitem{Osipov:2004} A.A.Osipov, H. Hansen, B.Hiller, Nucl.Phys. {\bf A745}, 81 (2004), hep-ph/0406112.
\bibitem{Eguchi:1975} T. Eguchi, Phys. Rev. {\bf D14}, 2755 (1975).
\bibitem{Reinhardt:1988} H. Reinhardt and R. Alkofer, Phys. Lett. B {\bf 207}, 482 (1988).
\bibitem{Nikolov:1996} E.N. Nikolov, W. Broniowski, C.V. Christov, G. Ripka, K. Goeke, Nucl. Phys. {\bf A608}, 411 (1966), hep-ph/9602274.
\bibitem{Osipov:2001} V. Bernard, A.H. Blin, B. Hiller, Y.P. Ivanov, A.A. Osipov, Ulf-G. Meißner, Annals Phys. {\bf 249}, 499 (1996); A.A. Osipov, B. Hiller, A.H. Blin, Phys.Lett. {\bf B475}, 324 (2000) ,hep-ph/9912404;
         A.A. Osipov and B. Hiller, Phys. Rev. D {\bf 62},
        114013 (2000), hep-ph/0007102; Phys. Rev. D {\bf 63}, 094009
        (2001), hep-ph/0012294, A.A. Osipov, M. Sampaio, B. Hiller, Nucl. Phys. A {\bf 703}, 378 (2001),
        hep-ph/0110285.
\bibitem{Dias2006}E.W. Dias, B. Hiller, A.L. Mota, M.C. Nemes, M. Sampaio, A.A. Osipov, Modern Physics Letters A {\bf 21},
339 (2006), hep-ph/0503245.
\bibitem{andre:1999} A.L. Mota, M.C. Nemes, B. Hiller, H. Walliser, Nucl. Phys. A {\bf 652}, 73 (1999), hep-ph/9901455.
\bibitem{Amendolia86} S.R. Amendolia et al., Phys. Lett. B {\bf 178}, 435
(1986).
\bibitem{Amendolia84} S.R. Amendolia et al., Phys. Lett. B {\bf 138},
454 (1984).
\end{thebibliography}
\end{document}